\documentclass[preprint,12pt,nonatbib]{elsarticle}   

\usepackage[american]{babel}
\usepackage{csquotes}                                
\usepackage[style=apa, backend=biber]{biblatex}
\DeclareLanguageMapping{american}{american-apa}
\usepackage{lmodern}
\usepackage{subcaption} 
\usepackage{amssymb}
\usepackage{amsmath,amsfonts,amssymb}
\usepackage{graphicx}
\usepackage{multirow}
\usepackage{lscape}
\usepackage{blindtext}
\usepackage{longtable}
\usepackage[colorlinks=true, allcolors=blue]{hyperref}
\usepackage{booktabs,caption}
\usepackage[flushleft]{threeparttable}
\usepackage{graphicx}
\usepackage{tablefootnote}
\usepackage{float}
\usepackage[margin=42mm]{geometry} 
\usepackage{mdframed}
\usepackage{amsmath}
\usepackage{lscape}
\usepackage{longtable}
\usepackage[utf8]{inputenc}
\usepackage{array}
\usepackage[T1]{fontenc}
\usepackage{tabularx}

\newcommand{\NCR}[1]{{\color{black}  #1}}

\journal{}

\begin{document}

\begin{frontmatter}

\title{The Feature Understandability Scale for Human-Centred Explainable AI: Assessing Tabular Feature Importance} 

\author[label1,label2]{Nicola Rossberg}
\author[label3,label4]{Bennett Kleinberg}
\author[label1,label2,label5]{Barry O'Sullivan}
\author[label1]{Luca Longo}
\author[label1,label2,label4]{Andrea Visentin}

\affiliation[label1]{organization={School of Computer Science and Information Technology, University College Cork},
            city={Cork},
            country={Republic of Ireland}}

\affiliation[label2]{organization={Taighde Éireann - Research Ireland Center for Research Training in Artificial Intelligence},
            addressline={University College Cork},
            city={Cork},
            country={Republic of Ireland}}

\affiliation[label3]{organization = {Department of Security and Crime Science, University College London, United Kingdom}}
            
\affiliation[label4]{organization = {Department of Methodology and Statistics, Tilburg University}, 
    city = {Tilburg},
    country = {The Netherlands}}

\affiliation[label5]{organization = {Insight RI Centre for Data Analytics, University College Cork, Ireland}}

\begin{abstract}
As artificial intelligence becomes increasingly pervasive and powerful, the ability to audit AI-based systems is growing in importance. However, explainability for artificial intelligence systems is not a one-size-fits-all solution; different target audiences have varying requirements and expectations for explanations. While various approaches to explainability have been proposed, most explainable artificial intelligence methods for tabular data focus on explaining the outputs of supervised machine learning models using the input features. However, a user's ability to understand an explanation depends on their understanding of such features. Therefore, it is in the best interest of the system designer to try to pre-select understandable features for producing a global explanation of an ML model. Unfortunately, no measure currently exists to assess the degree to which a user understands a given input feature. 
This work introduces two psychometrically validated scales that quantitatively seek to assess users' perceived understanding of tabular input features for supervised classification problems. Specifically, these scales, one for numerical and one for categorical data, each with two factors and comprising 8 and 9 items, aim to assign a score to each input feature, effectively producing a rank and allowing for the quantification of feature prioritisation. A confirmatory factor analysis demonstrates a strong relationship between such items and a good fit of the two-factor structure for each scale. This research presents a novel method for assessing understanding and outlines potential applications in the domain of explainable artificial intelligence.
\end{abstract}

\begin{graphicalabstract}
\includegraphics[width=\linewidth]{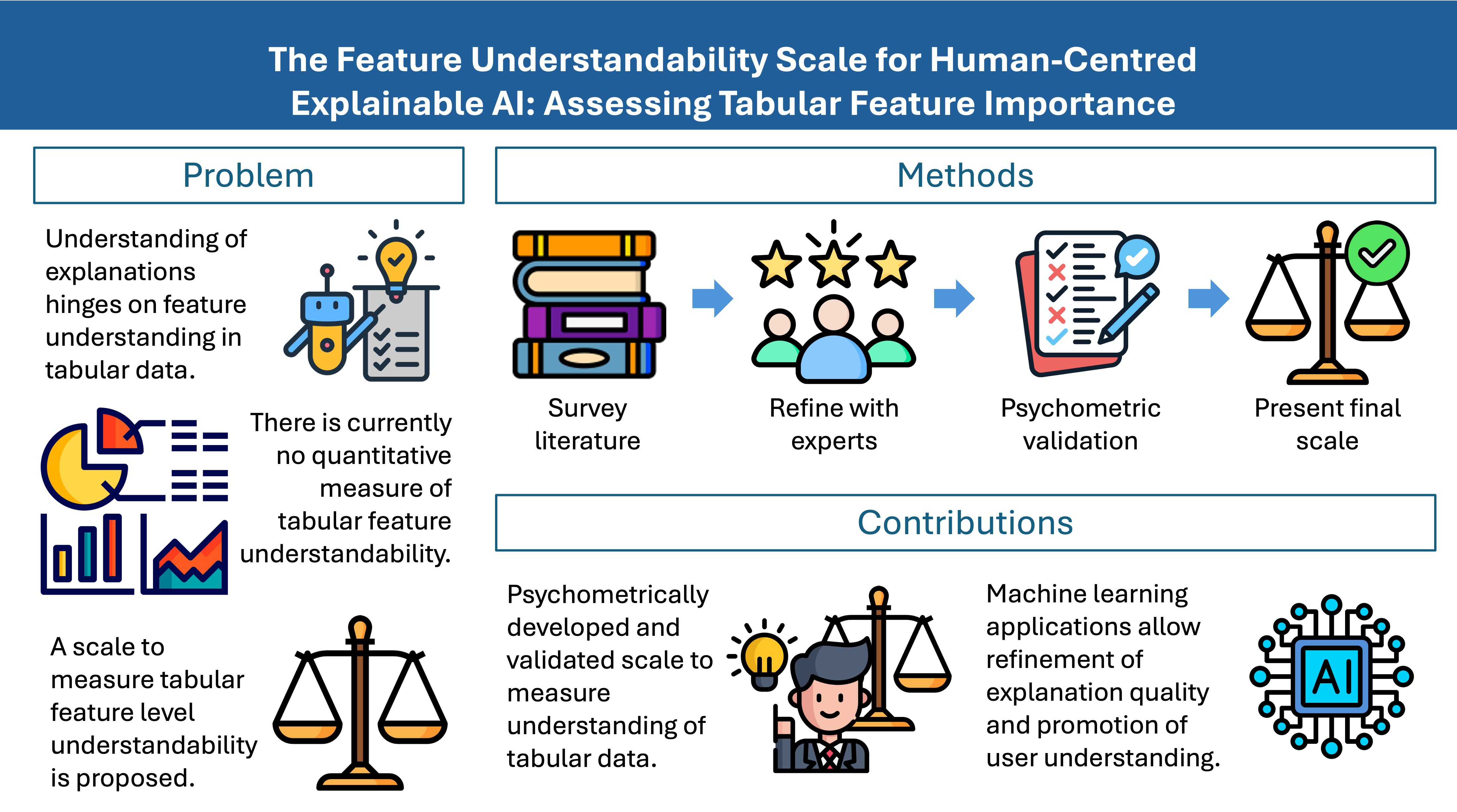}
\end{graphicalabstract}

\begin{highlights}
\item A validated scale assessing tabular feature understandability is presented. 
\item The scale quantifies the understandability of tabular features. 
\item The scale identifies understandable features, improving AI explanations.
\item This permits the co-optimisation of explainability and understandability.
\end{highlights}


\end{frontmatter}

\section{Introduction} \label{sec: Introduction}

In recent years, artificial intelligence (AI) has become increasingly pervasive across domains, prompting growing interest in leveraging its capabilities. However, many machine learning (ML) models remain black- or grey-boxes, limiting the extent to which their decision-making processes can be audited and understood \parencite{guidotti2018survey}. This opacity complicates the ability to ensure consistent and theoretically sound inferences. To address this, a range of explainability methods has been proposed to improve transparency, support auditability, and mitigate bias \parencite{wang2024roadmap}. These methods are particularly important in high-stakes domains such as medicine, where they contribute to both legal compliance and safe model deployment \parencite{longo2024explainable}.

Among eXplainable AI (xAI) approaches, textual explanations are widely used, especially for tabular data. These explanations describe model outputs in natural language by referencing input features and their values \parencite{vilone2021classification}. For example, a loan rejection may be explained by stating that the applicant’s `income' and `character' were too low. In such cases, user understanding depends on the understandability of the referenced features. While `income' is easily understood, `character' may be ambiguous, reducing the usefulness of the explanation. Prior work shows that understandable explanations are critical for fostering user trust and acceptance of ML outcomes \parencite{teso2023leveraging}. Interpretability-by-design approaches emphasise selecting features that are both relevant and comprehensible \parencite{chattopadhyay2023interpretable}. However, feature understandability is inherently subjective and dependent on the target audience, and no existing method systematically quantifies it.

This work addresses this gap by developing and validating a questionnaire-based scale to measure the understandability of features in tabular data, integrating psychometric methods into xAI research. The proposed Feature Understandability Scale is grounded in prior literature and validated following the ten-step framework by \textcite{carpenter2018ten}. The scale is refined through a multi-stage process, including exploratory and confirmatory factor analysis on a sample of 1,440 participants.

The scale measures the extent to which specific user groups understand tabular features, enabling explanations to be tailored accordingly. It is designed for deployment among target users, with collected scores informing model training via cost-sensitive feature selection to prioritise more understandable features. The unit of analysis is feature-level understanding, nested within participants and domains, and is based on self-reported perceptions rather than objective comprehension. While understanding may differ by domain and context, the distribution of the scale to a representative sample of a given user group will permit the assessment of the average self-assessed understanding of features within the population. By enabling the assignment of an “understandability” score to each feature, this work provides a novel tool for integrating user understanding into ML design. To our knowledge, this is the first scale to quantify feature-level understandability in tabular datasets. This study contributes to the interpretability-by-design literature.

To summarise, this research aims to fulfil four key goals:

\begin{enumerate}
    \item Propose and refine a scale measuring feature understandability through literature and expert feedback.
    \item Empirically validate the scale through large-scale data collection and exploratory factor analysis.
    \item Confirm and validate the final scale structure through confirmatory factor analysis.
    \item Proposing future avenues for the integration of feature-understandability scores for interpretability-by-design solutions.
\end{enumerate}

The design and validation of psychometrically sound measures hinge on proper scale development and reporting, which is frequently absent or inconsistent in the literature \parencite{carpenter2018ten, rottweiler2021measuring}. The disparities resulting from the lack of standards lead to questionable reliability and validity, as well as significant quality discrepancies between developed scales. To ensure high standards during the development of this study, our methodology is based on \textcite{carpenter2018ten}'s ten steps for scale development, which are detailed in Section \ref{sec: Methods}. 

The remainder of this manuscript is structured as follows. Section \ref{sec: Literature Review} summarises the known findings on xAI stakeholder taxonomies and needs. Based on such a review, an initial scale is proposed. Section \ref{sec: Methods} describes the methodology for the scale refinement and subsequent psychometric validation following the ten steps proposed by \textcite{carpenter2018ten}. The initial design and subsequent refinement through expert interviews, focus groups, and a pilot study, detailed in Section \ref{sec: Methods}, satisfy the first aim of the study. The subsequent exploratory and confirmatory factor analysis successfully met the aims of the second and third goals of this research. Section \ref{sec: Results} presents the outcomes of the design and validation steps, along with the scale alterations resulting from each step. Finally, Section \ref{sec: Discussion} analyses the most important findings of this study and presents avenues for applications and future research, satisfying the final objective of this work.

\section{Literature Review} \label{sec: Literature Review}
This section is divided into four parts. Section \ref{sec: Background} situates the research within the domain, Section \ref{subsec: Definitions} provides the relevant definitions, Section \ref{subsec: Theoretical Concepts} reviews the known stakeholder needs in xAI. and Section \ref{subsec: initial scale} describes the construction of the initial scale, which is presented as \textit{Draft 1}. All scale drafts can be found on GitHub\footnote{\url{https://github.com/ncrossberg/Feature-Understandability-Scale}}.

\subsection{Background} \label{sec: Background}

\subsubsection{Explainability Methods}

A wide array of xAI techniques has been proposed to address varying explanation needs. \textcite{wang2024roadmap} provides a comprehensive overview of the existing stakeholder groups, their respective explainability needs and the xAI methods suitable for meeting these needs. \textcite{saranya2023systematic} review the current state of the art of existing xAI models and application domains. In addition to general ad-hoc xAI techniques, domain-specific methods have been proposed, including the medical field \parencite{muhammad2024unveiling, contreras2025explainable, antoniadi2021current, band2023application}, with other domains including finance \parencite{weber2024applications} and transportation  \parencite{nwakanma2023explainable}. Ad-hoc explanations provide insight into the machine learning system's explanatory pathways and can be tailored, to some extent, to the needs of the outcome group, adjusting the level and detail of the explanation. However, the majority of explanatory methods for tabular data still hinge on references to feature inputs \parencite{vilone2021classification, sahakyan2021explainable} and their accessibility is limited to the user's ability to understand explanations. While current methods allow some adjustment to user groups, integrating understandability into machine learning model training would enable explanations to be created at the correct level by design.

\subsubsection{Explanation Evaluation Measures} Previous studies have evaluated the quality of explanations that were consulted during the development of the initial scale. The work by \textcite{holzinger2020measuring} proposes the system causability scale, which provides a quick, simple method for assessing the causability of an explanation interface or an explanation process. The system causability scale was based on the system usability scale originally proposed by \textcite{brooke1996sus}, which was designed to measure the usability of an industrial system in a `quick and dirty' way. As such, it is essential to note that these scales were designed with brevity rather than absolute validity in mind, which limits their applicability in the current context. In the same vein, \textcite{vilone2023development} proposed a 12-item questionnaire to evaluate the quality of explanations generated by xAI approaches. This research differs from these existing scales in several key aspects, addressing important stakeholder needs. On the one hand, it provides a measure of understandability at the feature level, thereby increasing measurement granularity. Additionally, the scale is vigorously psychometrically validated, permitting replication and improving measurement thoroughness and reliability. Finally, quantifying feature-level understandability enables a priori tailoring of explanations to the understanding of a given user base, which is not possible with the existing scales.

\subsubsection{Accuracy-Explainability Trade-Off}

The existence of an explainability-accuracy trade-off is a recurring topic in the literature. Some argue that more explainable models decrease accuracy because their classification algorithms are simpler \parencite{ hunsicker2025investigating, kabir2025review, assis2025performance}. Others believe that this trade-off is theoretically sound but not generally observed in practice, partially due to the lack of measurement instruments for interpretability \parencite{bell2022s, dziugaite2020enforcing, carrington2018measures}. In addition to general model interpretability, post-hoc xAI methods such as GradCAM, SHapley Additive exPlanations (SHAP) and LIME exist, which can generate explanations after model training \parencite{selvaraju2017grad, lundberg2020local2global, ribeiro2016should}. However, post-hoc approaches may be less faithful to the system, as they provide only approximations of classification processes.

\subsection{Definitions} \label{subsec: Definitions}

\NCR{
\paragraph{Understanding}
In the literature, the concept of `understanding' is more vague than that of explainability, despite definitions of explainability frequently hinging on the concept of understanding, with researchers defining explanations as `statements that make AI systems understandable for users' \parencite{brennen2020people, adadi2018peeking}. It is important to distinguish between technical interpretability, which is the transparency of the model’s internal logic and cognitive understanding. This distinction is relevant as an explanation may be technically complete but fail to promote understanding if it exceeds the user's cognitive load or fails to align with the social requirements of a "good" explanation, such as contrastiveness and selectivity \parencite{miller2019explanation, van2021evaluating}.

\paragraph{Feature Understandability} The overarching aim of the scale is to quantify 'feature understandability', defined as the degree to which users perceive that they can comprehend both what a feature means and how it may be related to the outcome variable of a given prediction task on tabular data. This can be broken down into two components: 1) comprehension of what the feature means and how it is measured, and 2) the perceived relevance and sensibility of the feature as an input for predicting the outcome. The second does not assess whether a feature is actually causal, but rather captures whether users understand why a feature would be used. Feature understandability is a feature-level construct assessed prior to model construction and is model and explanation-independent. It differs from explanation usefulness, trust and interpretability in its level of measurement (data, not explanation level) and timing (before model construction).

Understandability is user-relative and may be moderated by domain knowledge \parencite{pascual2014comprehension}. The scale, therefore, measures perceived understanding within a given target group, with the group-level mean scores indicating which features are likely to be understood most easily within the group. Feature understandability is, as such, not an expertise-independent property.
}

\subsection{Stakeholder Needs} \label{subsec: Theoretical Concepts}

To identify which aspects of explainability are important across user groups, the literature on stakeholder taxonomies and xAI needs was reviewed. Prior work consistently highlights substantial variation in users and their requirements \parencite{langer2021we, simkute2021explainability}, with multiple taxonomies emphasising that explainability must be tailored to the target audience \parencite{preece2018stakeholders, hind2019ted}. Accordingly, explanations should be adapted to users’ needs and levels of understanding.

A key finding is that stakeholders seek to understand not only model behaviour but also the input data driving predictions \parencite{hepenstal2020explainable, hoffman2023explainable, hong2020human, kim2024stakeholder}. Two aspects are particularly important: (1) understanding features in relation to the outcome, and (2) understanding their measurement and value ranges. For instance, \textcite{hepenstal2020explainable} found that stakeholders emphasise input data comprehension and measurement quality, motivating the initial subscales of \textit{Understanding} and \textit{Measurement}. Example items include “I can easily understand this feature,” and “The scale of the feature is easily understandable.” Similarly, \textcite{hoffman2023explainable} showed that users prioritise understanding inputs over model internals, motivating items such as “I require no support to understand the feature.” The importance of value ranges was further highlighted by \textcite{kim2024stakeholder}, leading to items such as “The scale of the feature is easily understandable,” and “I think I can verify the value of the feature with reasonable effort.”

A second finding is the importance of tailoring explanations to users’ knowledge and context. Explanations function as a knowledge-transfer process and must align with users’ backgrounds \parencite{ehsan2021expanding}. This is supported by \textcite{hoffman2023explainable}, where stakeholders emphasised accessibility and self-contained explanations (e.g., “Can a user, without training, figure out how to use the system within 10 min?”). Consequently, items capturing ease of understanding were included, such as “I require no support to understand the measuring scale of the feature.”

Finally, stakeholders exhibit divergent conceptions of explainability, necessitating alignment with user expectations \textcite{brennen2020people}. Desired properties include independent verification (“verifiable”, “self-service”), ease of comprehension (“understandable”), and insight into decision-making (“auditable”). This motivated the inclusion of a \textit{Faithfulness} subscale \parencite{vilone2021notions}, capturing perceived relevance and feature usage, with items such as “I understand how this feature is related to other features,” and “I think the feature should be included in the inputs”.

These findings align with broader concepts in the literature, including Selection and Simplicity \parencite{miller2017explainable}, Informativeness \parencite{lipton2018mythos}, and Explicitness \parencite{jaakkola2018towards}. Across all perspectives, \textit{Comprehensibility}, the clarity and accessibility of explanations, emerges as a fundamental requirement \parencite{dovsilovic2018explainable, askira1998knowledge}. Quantifying feature understandability is therefore essential for producing effective and user-centred explanations.

\subsection{Initial Scale} \label{subsec: initial scale}

The initial scale was drafted based on both the reviewed stakeholder needs and existing post-hoc explanation quality measures. The supplementary materials detail the source for each scale item. It is important to note that while some items were directly adapted from existing scales (e.g. Item 2, `I require no support to understand the feature.' adapted from \textcite{vilone2023development}'s `External support was required to understand the explanatory method, e.g., books, internet search, another person'), others were adapted, integrating the identified stakeholder needs. The postulated subscales were included for organisational purposes only, and items were presented independently during scale validation to permit factors to emerge empirically. A brief description of the item generation steps is provided below.

After postulating the initial subscales based on the review described in Section \ref{subsec: Theoretical Concepts}, existing explanation-quality scales were reviewed. A two-step process was used to identify and adapt candidate items. First, the suitability of items at the feature level was inspected; if items could logically function to assess feature quality, they were considered candidates. Subsequently, items were rephrased, changing explanations to feature references and adapting the remaining structure to make grammatical sense. Finally, items were reviewed by a second author to assess consensus regarding their inclusion. The works by \textcite{holzinger2020measuring, vilone2023development, liao2020questioning} were evaluated for this purpose. In addition to the adaptation of items, important considerations were taken from these scales and included in the items taken from the literature review. One example of this is the `I think that most people' phrasing from \textcite{holzinger2020measuring}. As the scale aims to cover attitudes across a given user cohort, this was adapted to `I expect others affected by this decision' and included in several items. Several items were formulated based on a single item and stakeholder need, on a different scale, to account for variations in language and permit later refinement. The initial FUS consisted of three subscales (Understanding, Measurement and Faithfulness) with the Measurement scale divided into numerical and categorical data. This division permits sufficient specificity when querying measurement levels for different feature types. A combined measurement category was considered, but the lack of specification regarding categories, ranges, etc., limited its efficacy considerably. The following describes each step taken to improve, verify, and test the drafted scale.

\section{Methods} \label{sec: Methods}

The scale refinement was based on the methods proposed by \textcite{carpenter2018ten}. However, it is important to note that this study differs from traditional scale validation due to the nature of understanding. Where the validation of scales in traditional psychometrics depends on variations in participants alone, such as individual changes in narcissism or depression, the current study considers variations in both participants and features, as understanding hinges on their interaction. As such, the sample size and the resulting analysis were adjusted to consider not only the item-to-participant ratio but the item-to-feature-to-participant ratio, marking a novel contribution to psychometric development.

\begin{figure}
    \centering
    \includegraphics[width=\linewidth]{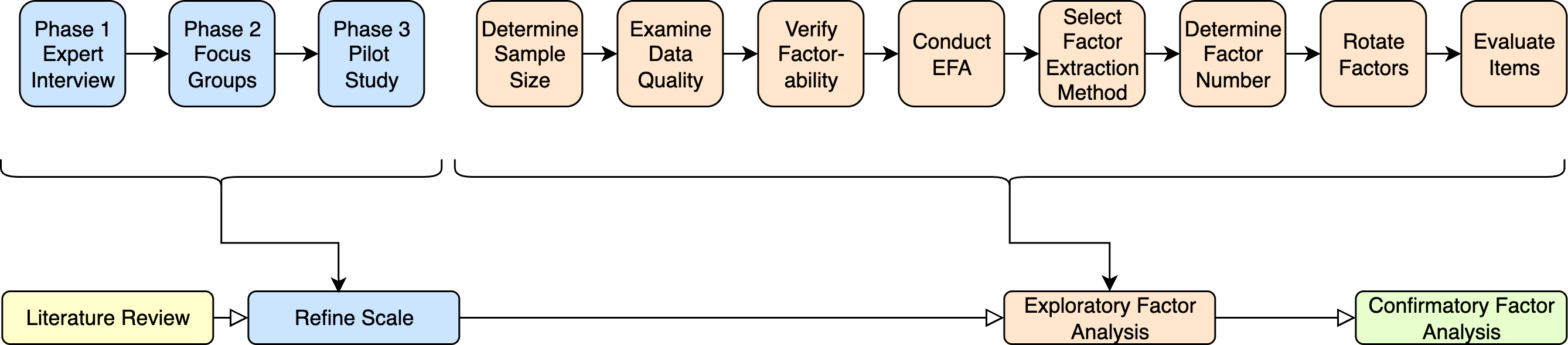}
    \caption{Workflow of the Psychometric Development of the Scale}
    \label{fig: Methods Workflow}
\end{figure}

Going forward, expert interviews were conducted to assess their understanding of the concept of explainability and to review the initial scale, identifying missing and redundant items. The integration of expert feedback is necessary for tailored considerations not apparent in the literature, as this is a novel approach to measuring understandability.
Next, two layperson focus groups were conducted to refine the phrasing and clarity of the items. This ensured that the level of items was comprehensible across user groups and not too specialised for laypersons to understand. Finally, a pilot study involving 89 participants was conducted to gather broader, anonymous feedback on the structure and clarity of the scale and to make any necessary final amendments before commencing large-scale data collection. Pilot studies provide further advantage over focus groups in that a larger participant pool can provide anonymous feedback, hence increasing the number of responses and reducing the impact of potential acquiescence bias. \NCR{Ethics approval for the expert interviews, focus groups and pilot study was granted on the 25th of September 2024 under Log Number 2024-166. Ethics permission for the subsequent Prolific-based data collection was granted on the 7th of July 2025 under Log Number 2024-166A2. All permissions were granted by the University College Cork Ethics Committee for Social Research before data collection commenced.} Below, each step is described in further detail. The complete workflow is illustrated in Figure \ref{fig: Methods Workflow}.

\subsection{Refining Scale}

After the initial draft, the scale was refined through expert interviews, focus groups and a pilot study. Both expert interviews and focus groups were analysed via inductive coding, and the scripts for the respective sessions can be found on GitHub. Themes were identified based on the lead researcher's interview notes and grouped accordingly. Disagreements between expert and focus group feedback were resolved in a two-step process. Where possible, majority agreement was favoured. If no majority existed, theoretical prioritisation based on the literature review of Section \ref{sec: Literature Review} was used to resolve disagreements.

\paragraph{Phase 1: Expert Interviews}

Expert interviews were conducted to assess item quality and the extent to which they capture the underlying construct. Six experts from academia and industry, all with at least a Master’s degree in fields such as computer science or psychometrics, were recruited. They were selected based on academic credentials and personal networks, and all worked on explainability, human-machine interaction or related topics at the time of the interview. Participants provided informed consent and were briefed on the study beforehand. The interviews had two parts: first, experts discussed their understanding of explainability; second, they evaluated the scale’s structure, suggested additions or removals, and reviewed item wording.

Three steps were taken to analyse the expert interviews. First, during the interviews, the lead researcher documented responses in writing. Particular care was taken to record the phrasing of additional items verbatim, and these verbatim suggestions were directly integrated into the resulting item set. Second, responses to the initial question on explainability and the understanding of explanations were analysed using an inductive coding approach. The researcher iteratively reviewed the notes, assigning codes to emerging concepts and refining these codes through repeated passes over the data to ensure internal consistency. Third, coded themes were grouped across interviews to identify recurring patterns. Based on these themes, additional items were formulated. As the analysis was conducted by a single researcher, particular attention was paid to coding consistency by revisiting the data multiple times and refining the coding scheme iteratively.

\paragraph{Phase 2: Focus Groups}

In addition to the feedback received during expert interviews, two focus groups were conducted, each with 2 and 3 laypeople, respectively \parencite{carpenter2018ten}. Small groups were chosen for this refinement step to allow an even discussion among all group members. As this step aimed to refine and clarify existing items, this setting enabled querying opinions from all participants. Laypeople were recruited to ensure the accessibility and clarity of language, as expert and literature-informed items may be overly convoluted. The focus group structure was similar to that of the expert interviews, where participants were first asked to share their general understanding of xAI before being debriefed about the research goal and presented with the designed scale. Based on the focus groups, item phrasings were adapted, and items were edited. The resulting scale was presented in Phase 3. 

\paragraph{Phase 3 - Pilot Study}

The final step of the initial scale design involved a pilot study implemented using Qualtrics (\url{https://www.qualtrics.com/}). This aimed to verify item structure, phrasing, and understanding through a larger, anonymous cohort. After completing the informed consent form, a brief explanation of the study's aim was provided, along with examples of features (both numerical and categorical), values, measurement scales, and endpoints for reference. Participants were then presented with each scale item and asked to provide any feedback, questions, or comments on clarity. Participants were recruited through convenience and snowball sampling, and no prior expertise in xAI or related fields was required, as feedback was sought only on item wording and structure.

\subsection{Exploratory Factor Analysis}
Exploratory Factor Analysis (EFA) is a statistical technique used to group data according to underlying latent variables \parencite{carpenter2018ten}. It rests on the assumption that the observed data can be divided into several related groups and by analysing inter-correlations in items, these groups or `factors' can be identified. It is frequently used in psychometrics to identify the latent constructs constituting a given measurement, as for instance grandiose and vulnerable narcissism functioning as factors of overall narcissism \parencite{sivanathan2024unified}). EFA is used to analyse and validate the constructed scale. The following describes the methodology employed in this study using EFA.

\paragraph{Determine Sample Size}

Adequate sample size is essential for statistical power, defined as the ability to reject a false null hypothesis while maintaining confidence in the results \parencite{anthoine2014sample}. While \textcite{worthington2006scale} suggests a minimum of 300 participants, other work emphasises item-to-participant ratios, with early recommendations of 1:5 or 1:10 \parencite{gorsuch1983} and more recent guidance suggesting up to 1:20 for robust solutions \parencite{costello2005best, rottweiler2021measuring}. In this study, sample size was determined by both item-to-participant and item-to-feature ratios, as understanding depends on both respondents and evaluated features. Nine features across three xAI domains (medicine, loan, hiring) were assessed, with 240 ratings per feature (2160 total). After splitting for EFA and CFA, this yielded ratios of 1:55 (numerical) and 1:47 (categorical), exceeding recommended thresholds.

We used a large-scale general population survey to collect data for the scale development. Participants were recruited via Prolific\footnote{Prolific connects researchers with participants, enabling the distribution of surveys and participant reimbursement. Link: \url{https://www.prolific.com/} } and were limited to participants residing in English-speaking countries (the United States, the United Kingdom, Ireland, Canada, Australia and New Zealand). Each participant completed an informed consent form during which they were debriefed regarding the purpose of the study. Subsequently, participants were asked to complete the questionnaire. Each participant was assigned three of the nine features and asked to rate them using the developed scale. Features were presented pseudo-randomly and with equal frequency, ensuring that the same number of ratings were collected for all features. The randomisation was implemented through Qualtrics' "randomiser" block, which permits the random presentation of a chosen number of features and permits the counter-balancing of presentation to ensure an even number of ratings for all features. Qualtrics' 'Display answers in a random order' option was used to shuffle the order in which scale items were presented and prevent order effects. The attention check "Attention Check: Please select Neutral for this item." was included with the scale items and presented at a random point for each feature. All items were measured on a 5-point Likert Scale, ranging from `strongly disagree' (1) to `strongly agree' (5). The Likert Scale anchors were the same for numerical and categorical items. Features were selected from xAI-relevant areas as identified by \textcite{islam2022systematic} and taken from domain-specific xAI research \parencite{rabie_el_kharoua_2024, mahmoud2019performance, lee2021patient, xu2024using, namdari_seerbc_2022, yasserh_loan_default}. The supplementary materials show the selected features and their intended meaning. The selection of features was based on the theoretical generalisation across all domains. However, empirical generalisation remains limited to the chosen areas of hiring decisions, medicine and loan applications.

\paragraph{Examine Data Quality}

To ensure the reliability of our findings, several pre-checks are implemented to test data quality before analysis. These checks ensure that low-quality or non-human responses are identified and excluded. Attention checks were implemented for each feature evaluation, asking participants to select the `neutral' option for one item. If individuals fail the attention check, their responses to this feature are discarded. The following metrics were consulted to assess data quality before analysis:

\begin{enumerate}
    \item Attention checks \parencite{abbey2017attention};
    \item Completion time (very fast time may indicate lack of focus) \parencite{rottweiler2021measuring};
    \item Missing Responses \parencite{carpenter2018ten}.
\end{enumerate}

Data preprocessing was conducted in Python, and the analysis was conducted in R using the `psych' package \parencite{psych}. The complete code used for analysis can be found on GitHub.

\paragraph{Verify the Factorability of the Data}

The next step in EFA is to check whether the data can be factored. Factorisation refers to the identification of underlying subscales that group the items. To verify that data can be factored, the following tests were inspected:

\begin{enumerate}
    \item Correlation Matrix - values of 0.30 or higher \parencite{carpenter2018ten}
    \item Bartlett's Chi-Square test with p $\leq 0.05$ \parencite{tabachnick2007using}
    \item Kaiser-Meyer-Olkin (KMO) of 0.60 or higher \parencite{tabachnick2007using}
\end{enumerate}

Based on these tests, the violating items were inspected, and deletion was considered if they were inconsistent with the overarching concept.

\paragraph{Conduct Exploratory Factor Analysis}

After cleaning and confirming factorability, EFA was conducted. It identifies factors of common variance in the data. The following steps were taken to run EFA for the construction and validation of the final \textit{Feature Understandability Scale}.

\paragraph{Select Factor Extraction Method}

Factor extraction identifies latent variables. There is limited information regarding the benefits of different extraction methods. While the sample size exceeds recommendations, it violates the assumption of multivariate normality and `weighted least squares' is selected for factor extraction. Weighted least squares is recommended for surveys with Likert-type scales \parencite{schreiber2021issues}

\paragraph{Determine Factor Retention Criteria}

After factor extraction, the number of factors to retain was determined. Generally, it is recommended to either choose one method in advance of data analysis or examine convergence between different methods. Three criteria were used to determine the number of factors to retain, and their convergence is examined. The three chosen criteria are a `parallel analysis with 99\% threshold', `eigenvalue greater than 1' and the `scree elbow test,' \parencite{carpenter2018ten}. 

\paragraph{Rotate Factors}

Rotation transforms the factor-loading matrix to simplify its structure and facilitate the identification of item-factor relationships. Rotation can be conducted in either an oblique or an orthogonal fashion. Oblique rotation is recommended in combination with weighted least squares factor analysis, and `Promax' is chosen, as it begins with an orthogonal solution and then transforms it to an oblique solution, making it more robust \parencite{thompson2004exploratory, schreiber2021issues}.

\paragraph{Evaluate Items Based on A Priori Criteria}

After these analyses, the number of items should be reduced. Ideally, the initial scale should include 3 times as many items as necessary. The following criteria were used to reduce items:
\begin{enumerate}
    \item Minimum factor item loading $>0.32$ \parencite{carpenter2018ten};
    \item Items cross-loading $>0.32$ on other factors are excluded \parencite{costello2005best, taherdoost2014exploratory};
    \item Minimum number of 3 items per factor
    \parencite{taherdoost2014exploratory};
    \item Examination of theoretical convergence of items \parencite{carpenter2018ten};
    \item Parsimony to minimise the redundancy of wording or meaning across items \parencite{carpenter2018ten}. 
\end{enumerate}

\subsection{Confirmatory Factor Analysis}
In addition to the 10 steps proposed by \textcite{carpenter2018ten}, Confirmatory Factor Analysis (CFA) ensures that the scale developed during EFA can be identified in new data, further validating its structure. The methodology for CFA in this study is based on \textcite{rottweiler2021measuring}.

\paragraph{Participants}
Data collection for CFA was conducted simultaneously with EFA, and the data were split evenly between the two procedures. For each feature, 240 ratings were collected, resulting in a total of 2,160 ratings for CFA. The methodology was the same as that of EFA, with respondents recruited via Prolific, completing an informed consent form, and rating three features using the developed scale. 

\paragraph{Procedure}

CFA aims to verify whether the factor structure identified during EFA remains consistent in new data. The data were cleaned in Python and analysed in R using the packages `Lavaan' and `SemTools' \parencite{lavaan, semtools}. After completing the analysis, the fit indices were examined and evaluated against the following standards put forward by \textcite{hu1999cutoff}:

\begin{enumerate}
    \item Chi-Squared ($\chi^2$) non-significance with $p > 0.05$;
    \item Comparative Fit Index (CFI) $\geq 0.9$;
    \item Tucker-Lewis Index (TLI) $\geq 0.9$;
    \item Root Mean Square Error of Approximation (RMSEA) $\leq .08$;
    \item Standardised Root Mean Square Residual (SRMR) $\leq 0.08$.
\end{enumerate}

The data displayed a skewed distribution, and a `Weighted Least Squares Means and Variance Adjusted' estimator with ordered categories was used to account for this.

\section{Results} \label{sec: Results}

\begin{figure}[!ht]
    \centering
    \includegraphics[width=1.\linewidth]{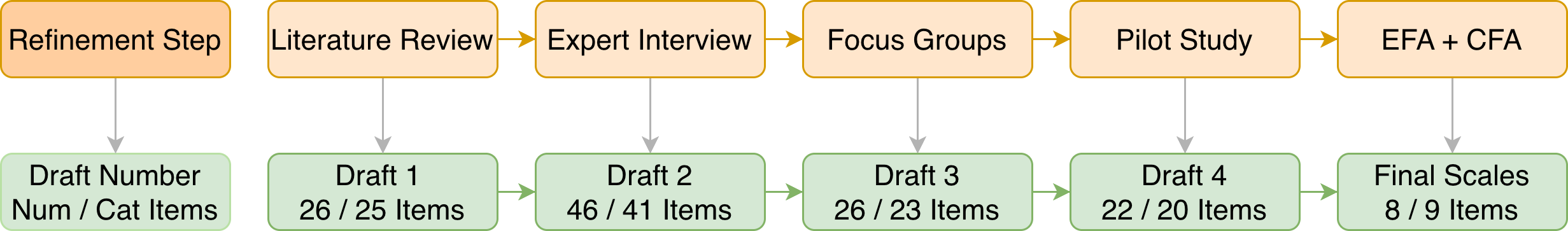}
    \caption{Flowchart detailing the outcomes of each refinement step. Item numbers are structured as numerical/categorical.}
    \label{fig: Flowchart_Drafts}
\end{figure}

Below are the outcomes of each refinement step. Figure \ref{fig: Flowchart_Drafts} summarises the draft and item number resulting from each step. The corresponding drafts are available on GitHub.

\NCR{The full analysis of themes and precise scale changes can be found in the supplementary materials.}

\subsection{Refine Scale: Expert Interviews}

The experts were posed several questions regarding their understanding of explainability, what they wish to gain from explainability, and their opinions on user needs for explanations. The full script for the semi-structured interviews is available on GitHub. The primary goal of the interviews was to gain a deeper understanding of experts' opinions and perspectives on explainability, and to solicit their thoughts on the developed scale. Three core topics were extracted from the responses, and new items were synthesised on their basis. Each topic is detailed in the supplementary materials, and examples of the extracted items are provided. It is important to note that several similarly phrased items are included, as this allows the better-functioning item to be selected during quantitative analysis.

The experts also recommended adding definitions at the top of the scale to clarify certain words, including `feature', `value', `outcome' and `scale'. These definitions will be provided to users before scale completion to ensure consistent understanding between users and minimise discrepancies caused by subjective interpretation of phrases.

\subsection{Refine Scale: Focus Groups} \label{sec: Focus Groups}

The first concept which arose from the discussion about general explainability was that of bias and fairness. One participant stated that subjects should rate the extent to which they believe the feature to be a fair way of measuring the underlying concept and whether the feature may introduce bias into the system. The argument behind this idea was that subjects should be able to attest to problematic features, such as postcodes for mortgage predictions, and that this should be taken into consideration during system design. However, while this is a valid consideration, it is not related to the subjects' understanding of the feature and, as such, is not included in the current scale.

The second consideration was drawn from the research area of counterfactuals, aiming to probe the participants' understanding of how different values of the feature may impact the outcome. It was argued that, as a component of understanding, participants should not only be aware of which features impacted the outcome but also grasp how changes in the values of a given feature may change or affect the outcome. An example of this is that participants not only understand that debt affects the likelihood of being granted a mortgage, but also that the model is more likely to reject an application as debt increases. This is introduced into the scale under the `Understanding' subscale using the following two items. Note that the alternative phrasings are introduced to allow analysis of item functioning, and only one phrasing will be maintained in the final scale. 

\begin{enumerate}
    \item I know how a change in the feature's value would impact the outcome.
    \item I know how a change in the feature's value would change the outcome
\end{enumerate}

The final adjusted scale can be found on GitHub labelled \textit{Draft 3}. In addition to the changes in phrasing, several definitions were added to the scale to clarify concepts and ensure consistent understanding among participants. After implementing the focus group feedback and revising each item individually for parsimony, 32 items remained.

\subsection{Refine Scale: Pilot Study} \label{sec: Pilot Study}
The final step in constructing the scale was a pilot study to gather comprehensive feedback on the phrasing and structure of each item. In the pilot study, participants were asked whether each item was understandable and if they had suggestions for rephrasing or clarification. A total of 89 responses from lay individuals recruited via convenience sampling were collected and qualitatively analysed. Based on the responses, several items were rephrased for clarity, and three items were removed. The most common concern was item length and vocabulary complexity, and simpler phrasings and more accessible synonyms were utilised in the revised scale. After these revisions, the final scale consisted of 28 items and is presented as \textit{Draft 4} on GitHub.

\subsection{Exploratory Factor Analysis} \label{sec: EFA}

Data were collected on August 28th 2025, using the platform `Prolific' in conjunction with `Qualtrics'. The sample consisted of adults aged 18 or older residing in the United Kingdom, the United States of America, Ireland, Canada, Australia, or New Zealand. The compensation rate was set at 9 pounds per hour in line with platform recommendations, and additional payments were made after study completion to satisfy Irish minimum wage requirements. A total of 1440 participants were recruited. The sample was split participant-wise, with 720 participants used for EFA and 720 retained for CFA, stratified by feature type (numerical or categorical) to ensure equal distribution between EFA and CFA. This resulted in 2160 ratings for EFA and CFA, respectively. Each participant rated three of the nine features of the scale shown in \textit{Draft 4} of the scales found in the additional material, and ratings were randomly assigned and stratified. For each of the nine features, 480 responses were received (240 for EFA and 240 for CFA). \NCR{The final demographics per feature are reported in Table \ref{tab: demographics}.}

\begin{table}[!ht]
\centering
\caption{Feature-wise demographics across both EFA and CFA.}
\label{tab: demographics}
\resizebox{\columnwidth}{!}{%
\begin{tabular}{@{}lcccccccccc@{}}
\toprule
\textbf{Feature} & \textbf{Min Age} & \textbf{Max Age} & \textbf{Mean Age} & \textbf{\% Female} & \textbf{USA} & \textbf{UK} & \textbf{Canada} & \textbf{Australia} & \textbf{Ireland} & \textbf{New Zealand} \\ \midrule
Recruitment Strategy & 18 & 79 & 43.34 & 49.1 & 20.6\% & 72.7\% & 3.7\% & 2.2\% & 0.6\% & 0.2\% \\
Degree & 19 & 79 & 43.34 & 49.1 & 18.9\% & 75.4\% & 2.5\% & 2.0\% & 0.4\% & 0.8\% \\
Neoplasm Stage & 19 & 77 & 43.28 & 50.6 & 18.9\% & 73.9\% & 2.5\% & 2.5\% & 1.7\% & 0.4\% \\
Business or Commercial & 19 & 78 & 42.58 & 47.5 & 22.4\% & 71.2\% & 1.9\% & 2.8\% & 1.3\% & 0.4\% \\
Personality Score & 18 & 79 & 42.6 & 49.0 & 20.3\% & 73.3\% & 2.3\% & 2.3\% & 1.3\% & 0.4\% \\
BMI & 19 & 79 & 42.6 & 47.7 & 19.6\% & 73.1\% & 2.3\% & 2.5\% & 2.3\% & 0.2\% \\
Total Cholesterol & 19 & 78 & 43.07 & 46.2 & 17.4\% & 72.9\% & 4.5\% & 3.9\% & 0.8\% & 0.4\% \\
Credit Score & 18 & 78 & 42.75 & 52.0 & 20.3\% & 70.9\% & 4.7\% & 2.7\% & 0.8\% & 0.6\% \\
Debt & 19 & 79 & 42.04 & 42.6 & 21.8\% & 72.0\% & 1.9\% & 3.0\% & 0.6\% & 0.6\% \\ \bottomrule
\end{tabular}%
}
\end{table}

\subsubsection{Examine Data Quality}

Data quality was examined based on built-in attention checks, completion time, and missing responses. Attention checks were conducted at the feature level, and 19 responses were removed due to failed attention checks. The mean response time was 9.5 minutes (SD = 5.2). Response time patterns were examined using a box-and-whiskers plot, and no responses were deleted based on response time. There were no missing responses. Missing responses should be deleted in a feature-wise fashion if present. After preprocessing, a total of 2141 responses remained, comprising 1199 ratings for the numerical features (1:55 item-to-response ratio) and 942 for the categorical features (1:47 item-to-response ratio), which is sufficient for analysis.

\subsubsection{Verify the Factorability of the Data}

The factorability of the data was verified using Bartlett's test of sphericity, KMO assessment and correlation analysis. Because the scale varies depending on whether it is used for numerical or categorical features, these analyses were conducted separately, and two final scales are presented. Multivariate normality of the data was tested using Mardia's test. Because the data were skewed, weighted least squares was used for the exploratory factor analysis.

As the data are ordinal, polychoric correlations are used for the exploratory factor analysis \parencite{barendse2015using, schreiber2021issues, muthen1985comparison}. For the initial numerical scale, the Barlett's Chi-Square $\chi\textsuperscript{2} (231)$ = 22243.7, p $\leq 0.001$, and the KMO = 0.94, and for the categorical scale, the Barlett's Chi-Square test $\chi\textsuperscript{2} (190)$ = 16510.06, p $\leq 0.001$, and the KMO = 0.97. This indicates excellent common variance and multivariate normality in the joint distribution of the items \parencite{kaiser1974index}. Item-item and item-total correlations were computed to assess whether items measured the same construct and check for highly correlated items. The full correlation matrices are available in the supplementary materials. \textcite{carpenter2018ten} recommends deleting items that correlate below 0.30 if it makes theoretical sense to do so. No items correlated below 0.3 for either dataset. These results indicated that factoring is possible, and analysis moved forward to the initial EFA.

\subsubsection{Conduct EFA}

The initial factor analysis was conducted with weighted least squares for factor extraction, the `Promax' rotation, and polychoric correlations. Based on a parallel analysis with a 99\% threshold, the scree elbow plot, and eigenvalues greater than one, three factors were retained for the numerical scale and two factors for the categorical scale. The outcome of the parallel analysis and the initial factor assignments are shown in the supplementary materials. Factor loadings were examined to assign items to the retained factors, and item-factor correlations were recomputed.

\subsubsection{Evaluate Items Based on A Priori Criteria}

After computing the initial sub-scale structures for the two scales, the included items were evaluated against a priori criteria. Items were removed if they had an item-factor loading $< 0.32$ (this removed two items from the numerical scale and none from the categorical scale). If items cross-loaded $>0.32$ on two factors, they were removed, leading to the deletion of no items from the numerical scale and three from the categorical scale. Factors with fewer than three items were then discarded. Finally, based on a qualitative examination of the theoretical convergence of items and the parsimony of wording and meaning across items, subscales were reduced. The final scales had 2 factors each, with 8 (numerical) and 9 items (categorical), respectively. 

\subsubsection{Present Results}

After item reduction, the EFA was re-run using weighted least squares factoring, `Promax' rotation and polychoric correlation. The analysis revealed a two-factor solution for both scales, comprising 8 items for the numerical scale and 9 items for the categorical scale. No cross-loadings $>0.32$ or weak loadings $<0.32$ remained for either scale. For the numerical scale, the first factor consists of 5 items and the second of 3. For the categorical scale, the first factor comprised 6 items, and the second, 3. The sum of squared loadings refers to the factor's variances after their extraction \parencite{tavakol2020factor}. For the numerical scale, the sums of squared loadings were 3.32 (41.5\%) and 2.46 (30.8\%), explaining a total of 72.3\% of the variance. The sum of the squared loadings for the categorical scale was 4.07 and 2.15 for each factor, respectively. This accounts for 45.2\% and 23.9\% of the variance, explaining 69.1\% of the total variance.

To assess scale homogeneity and ensure consistency between items measuring the same construct, polychoric item-total correlations were computed \parencite{zijlmans2018methods}. In the literature, a value of $\geq 0.30$ for each item-total correlation is considered sufficient; however, values between 0.30 and 0.70 are recommended. \parencite{de2013surveys}. For the numerical scale, item-total correlations ranged from 0.64 to 0.76, and for the categorical scale, from 0.64 to 0.82. This indicates that both scales have good homogeneity and inter-item consistency. Additionally, communalities for all items were assessed. Communalities indicate the proportion of each item's variance explained by the factors. Satisfactory communalities range from 0.40 to 0.70 and are considered good when above 0.7. For the numerical scale, communalities ranged from 0.57 to 0.82 with a mean of 0.72. For the categorical scale, communalities ranged from 0.63 to 0.80 with a mean of 0.71. As such, both scales met or exceeded the requirements for the proportion of item variance explained by the factors.

Finally, to ensure that no redundancy remained in the retained items, inter-item correlations were assessed \parencite{rottweiler2021measuring}. The full item-item correlation tables are available in the supplementary materials. Ideally, correlations should range from 0.20 to 0.50; items with correlations below 0.20 or above 0.80 should be considered for deletion. The item correlations ranged from 0.31 to 0.82 for the numerical scale and from 0.39 to 0.78 for the categorical scale. The average item-item correlation was 0.59 for the numerical scale and 0.64 for the categorical scale. To assess the internal consistency of the scales, McDonald's omega was computed, which is considered more robust than Cohen's alpha \parencite{hayes2020use}. The results, alongside the correlations of the factors, are presented in Table \ref{tab: EFA_metrics}. Both scales demonstrated very good internal consistency and moderate to strong factor correlations. The final scales, alongside factor loadings and polychoric item-total correlations, are presented in Tables \ref{tab: numerical_scale} and \ref{tab: categorical_scale}. It is interesting to note that for both scales, Factor 2, `Feature-Outcome Relation', consists of the same items. Only factor 1 varies as a function of the differences in variable conceptualisation (numerical or categorical). 

\renewcommand{\arraystretch}{1.2}
\begin{table}[!ht]
\centering
\caption{Metrics for the final scale after EFA.}
\label{tab: EFA_metrics}
\resizebox{0.5\columnwidth}{!}{%
\begin{tabular}{lcccc}
\hline
 & Factors & Mc Donald's $\omega$ & F1 & F2 \\ \hline
\multirow{2}{*}{Numerical} & F1 & 0.91 & 1 &  \\
 & F2 & 0.93 & 0.53 & 1 \\
\multirow{2}{*}{Categorical} & F1 & 0.94 & 1 &  \\
 & F2 & 0.88 & 0.66 & 1 \\ \hline
\end{tabular}%
}
\end{table}

The Feature Understandability Scale is split into two scales that measure understanding of numerical and categorical items, respectively. Both scales are two-factor scales. For the numerical scale, the first factor consists of 5 items and the second of 3. The categorical scale's first factor comprises 6 items, and the second comprises 3 items. The factors for both scales are named according to the same convention because of their similar item groupings. Names were chosen based on an inspection of the included items and the nomenclature of the initial scale. Factor 1, `Understanding and Measurement', contains items that measure the individual's ability to understand the feature itself as well as the method used to assess it. Factor 2, `Feature-Outcome Relation', comprises items assessing the respondent's perceptions of whether the feature is a sensible, relevant input for the model's prediction. The two scales are subsamples of the originally postulated subscales (Understanding, Measurement and Faithfulness). The naming conventions were adapted to align with the new structure. The first dimension, 'Understanding and Measurement', combines items of the first two subscales. This combination aligns with expectations, given that the understanding and measurement needs were based on the same research and hence are likely related \parencite{hepenstal2020explainable, hoffman2023explainable, kim2024stakeholder}. The 'Feature-Outcome Relation' dimension is a subsample of the 'Faithfulness' subscale, focusing solely on the relationship between the input data and the final outcome, and hence was renamed accordingly. \NCR{It is important to note that items referring to a feature `influencing' the outcome are intended to refer to the feature's role in the model's prediction, not a causal relationship. This factor reflects that understanding a feature in a prediction task requires more than knowing what it measures, but also why it may be associated with the outcome. Prior work showed that the understanding of both the input and its relationship to the output is crucial for explanation quality \cite{hepenstal2020explainable, hoffman2023explainable, hong2020human, kim2024stakeholder}. This is reflected in the two-factor structure of the scale.} Within each scale, the factors showed a moderate to strong positive correlation \parencite{akoglu2018user}. The factors for the numerical scale were correlated with $\phi = 0.53$, and for the categorical scale with $\phi = 0.66$.

\begin{table}[!ht]
\centering
\footnotesize
\setlength{\tabcolsep}{4pt}
\renewcommand{\arraystretch}{1.2}
\caption{Feature Understandability Scale for Numerical Items. Factor loadings (bootstrap standard errors in parentheses, 1,000 resamples) are displayed alongside item descriptive statistics (mean and standard deviation), polychoric item-total correlation ($\rho$) and communalities ($h^2$).}
\label{tab: numerical_scale}
\begin{tabularx}{\textwidth}{@{}l X c c c c c@{}}
\toprule
 & & \multicolumn{2}{c}{Loading (SE)} & & & \\ \cmidrule(lr){3-4}
N & Item & Factor 1 & Factor 2 & M (SD) & $\rho$ & $h^2$ \\ \midrule
\multicolumn{7}{@{}l}{\textit{Understanding \& Measurement}} \\
1 & I can understand the scale (units) of the feature. & 0.69 (0.031) & & 3.73 (1.03) & 0.70 & 0.57 \\
2 & I can easily understand if a given value of the feature is high or low. & 0.74 (0.030) & & 3.93 (0.92) & 0.68 & 0.58 \\
3 & I know what this feature measures. & 0.90 (0.021) & & 3.96 (0.86) & 0.75 & 0.78 \\
4 & I know what this feature represents. & 0.88 (0.022) & & 4.03 (0.83) & 0.76 & 0.78 \\
5 & I think that I can easily access a definition of the feature. & 0.84 (0.024) & & 3.92 (0.92) & 0.66 & 0.64 \\ \addlinespace
\multicolumn{7}{@{}l}{\textit{Feature-Outcome Relation}} \\
6 & In my opinion, the feature should be used to predict the outcome. & & 0.92 (0.016) & 3.51 (1.10) & 0.64 & 0.81 \\
7 & I think that the feature is important for the outcome. & & 0.88 (0.018) & 3.80 (0.99) & 0.69 & 0.82 \\
8 & I think it is fair that the feature influences the outcome. & & 0.90 (0.016) & 3.56 (1.10) & 0.66 & 0.81 \\
\bottomrule
\end{tabularx}
\end{table}

\begin{table}[!ht]
\centering
\footnotesize
\setlength{\tabcolsep}{4pt}
\renewcommand{\arraystretch}{1.2}
\caption{Feature Understandability Scale for Categorical Items. Factor loadings (bootstrap standard errors in parentheses, 1,000 resamples) are displayed alongside item descriptive statistics (mean and standard deviation), polychoric item-total correlation ($\rho$) and communalities ($h^2$).}
\label{tab: categorical_scale}
\begin{tabularx}{\textwidth}{@{}l X c c c c c@{}}
\toprule
 & & \multicolumn{2}{c}{Loading (SE)} & & & \\ \cmidrule(lr){3-4}
N & Item & Factor 1 & Factor 2 & M (SD) & $\rho$ & $h^2$ \\ \midrule
\multicolumn{7}{@{}l}{\textit{Understanding \& Measurement}} \\
1 & I can easily understand how the categories were assessed. & 0.91 (0.028) & & 3.30 (1.22) & 0.82 & 0.80 \\
2 & I can easily understand the order of categories. & 0.90 (0.027) & & 3.49 (1.14) & 0.77 & 0.74 \\
3 & I think it is feasible for me to verify the specific category of the feature. & 0.69 (0.035) & & 3.49 (1.08) & 0.78 & 0.67 \\
4 & I understand all possible values of the categorical feature. & 0.84 (0.033) & & 3.34 (1.20) & 0.71 & 0.63 \\
5 & I know what this feature represents. & 0.82 (0.030) & & 3.56 (1.16) & 0.81 & 0.75 \\
6 & I require no support to understand the feature. & 0.77 (0.034) & & 3.62 (1.12) & 0.79 & 0.70 \\ \addlinespace
\multicolumn{7}{@{}l}{\textit{Feature-Outcome Relation}} \\
7 & In my opinion, the feature should be used to predict the outcome. & & 0.82 (0.037) & 3.42 (1.01) & 0.66 & 0.68 \\
8 & I think that the feature is important for the outcome. & & 0.79 (0.039) & 3.80 (0.92) & 0.64 & 0.65 \\
9 & I think it is fair that the feature influences the outcome. & & 0.90 (0.039) & 3.54 (0.98) & 0.70 & 0.80 \\
\bottomrule
\end{tabularx}
\end{table}

\subsection{Confirmatory Factor Analysis}

The CFA aimed to confirm the factor structure identified during the EFA. The following statistics were tested: The chi-squared statistic was assessed, the CFI $\geq 0.90$, TLI $\geq 0.90$,
RMSEA $\leq .08$, SRMR $\leq .08$ \parencite{hu1999cutoff}. 

\subsubsection{Data Cleaning}

Before running the CFA, the data were cleaned. The procedure described in Section \ref{sec: EFA} was followed, using the second part of the data (n = 720). Responders missed attention checks for a total of 11 items, for which the responses were set to N/A and removed item-wise. This resulted in a total of 2,149 ratings (1,198 for the numerical scale and 951 for the categorical scale). No answers were discarded due to slow response time or incomplete answers.

\subsubsection{Results CFA}

\begin{figure}[ht!]
    \centering
    \includegraphics[width=\linewidth]{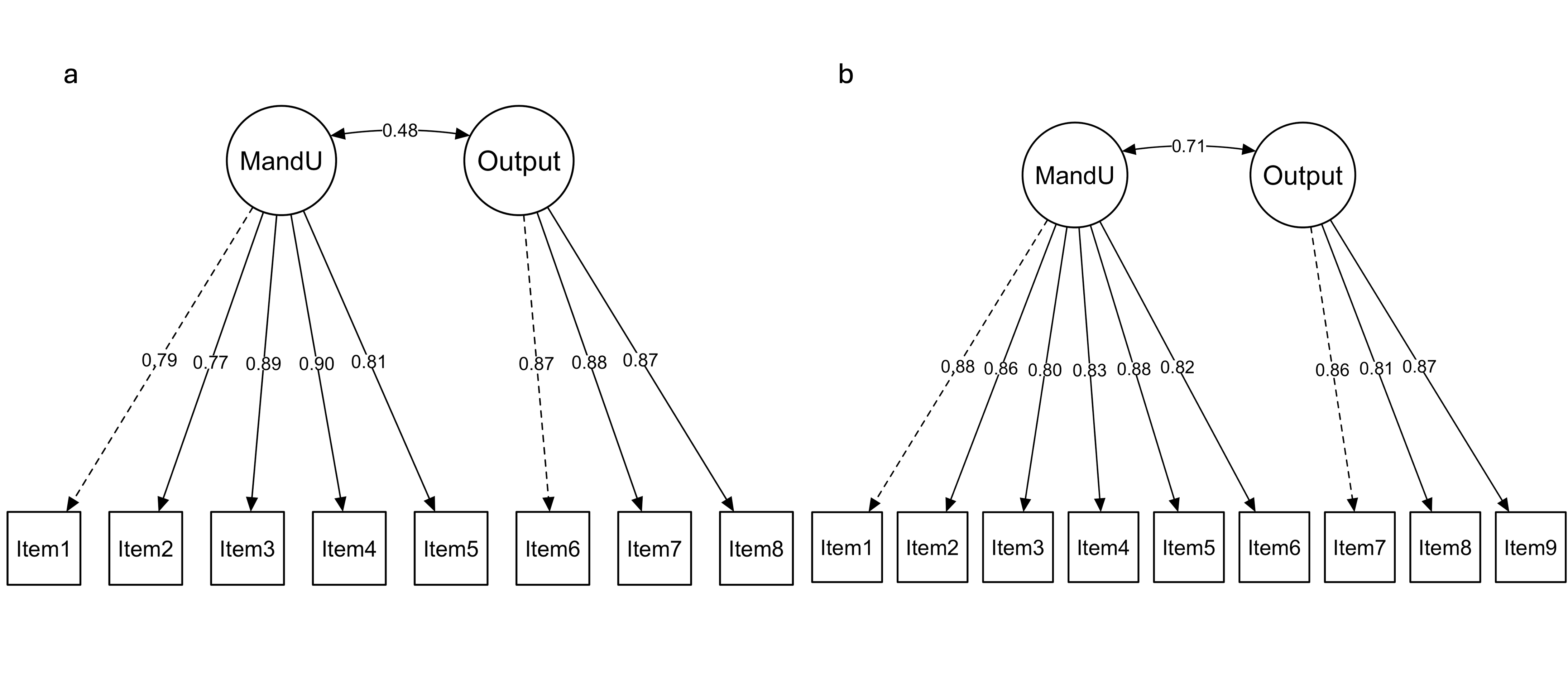}
    \caption{Result of the CFA for the a) Numerical Scale b) Categorical Scale. The Abbreviations are as follows: MandU (Understanding and Measurement) and Output (Feature-Outcome Relation).}
    \label{fig: Num_CFA}
\end{figure}

The CFA was run with a robust estimator, as both the numerical and categorical data showed skewness. The results of the confirmatory factor analysis are displayed in Table \ref{tab: CFA_Results}. Overall, the numerical model's fit is acceptable, and the categorical model demonstrates excellent fit. The factor loadings for the scales ranged from 0.77 to 0.90 for the numerical scale and 0.80 to 0.88 for the categorical scale, demonstrating strong relationships. These results indicate a good fit of the two-factor scales, and the final CFA models are shown in Figure \ref{fig: Num_CFA}.

Based on the methodology in \textcite{rottweiler2021measuring}, the current 2-factor scales were compared to a one-factor solution, where all items loaded on a single factor. This was done to ensure that the chosen two-factor structure was the best-fitting model. For both scales, the one-factor fit was extremely poor, and the results are shown in Table \ref{tab: CFA_Results}. To establish whether the one-factor model functioned significantly worse than the two-factor model, the chi-squared indices were compared between the two models. Since the data are skewed, a Satorra-Bentler scaled chi-square difference test was implemented for this comparison. For the numerical model, the one-factor solution was found to be significantly worse with $\chi\textsuperscript{2}(1)$ = 2468.67 $p \leq 0.0001$. For the categorical scale, the one-factor model also performed significantly worse with $\chi\textsuperscript{2}(1)$ = 1165.622 $p \leq 0.0001$. In addition, the change in the remaining metrics was computed and is shown in Table \ref{tab: CFA_Results}. Based on these results, the one-factor solution is rejected, and the two-factor scales are accepted. \NCR{It is important to note that in the two-factor solution, the RMSEA also exceeds the recommended cut-off of 0.08 on the numerical scale. A high RMSEA frequently occurs in models with small degrees of freedom, and \cite{kenny2015performance} suggest caution in its interpretation in these cases. \cite{shi2020assessing} furthermore found that RMSEA is highly likely to reject good model fit in data with five categories (such as found in this study) and large sample sizes and instead recommend the usage of SRMR, as a less biased estimate. Given that the SRMR is below 0.04 for both models, and the CFI and TLI estimates are furthermore above 0.98, it is likely that the high RMSEA is an artefact of the small degrees of freedom as well as the data structure. While it is still important to report its values, both for completion and transparency, the two-factor scale is found to perform well overall and is retained over the one-factor solution. As such, the \textit{Feature Understandability Scale} is best constructed with two factors, with 8 items for numerical features and 9 for categorical features. The two factors represent (1) the understanding and measurement of the feature and (2) the relation of the feature to the predicted output.}

\renewcommand{\arraystretch}{1.2}
\begin{table}[]
\centering
\caption{Results of the 2-factor and 1-factor analysis of the CFA alongside the changes in performance metrics.}
\label{tab: CFA_Results}
\resizebox{\columnwidth}{!}{%
\begin{tabular}{lllllllll}
\hline
 &  & $\chi^2$ & $\chi^2$ df & $\chi^2$ p & CFI & TLI & RMSEA {[}90\%CI{]} & SRMR \\ \hline
\multirow{2}{*}{Numerical} & 2-F & 210.957 & 19 & $p \leq 0.001$ & 0.989 & 0.984 & 0.092 {[}0.081, 0.103{]} & 0.034 \\
 & 1-F & 2468.67 & 20 & $p \leq 0.001$ & 0.865 & 0.810 & 0.320 {[}0.309, 0.331{]} & 0.185 \\
$\Delta$ &  &  &  &  & 0.124 & 0.174 & 0.228 & 0.151 \\
\multirow{2}{*}{Categorical} & 2-F & 181.624 & 26 & $p \leq 0.001$ & 0.991 & 0.988 & 0.079 {[}0.069, 0.090{]} & 0.028 \\
 & 1-F & 1165.622 & 27 & $p \leq 0.001$ & 0.935 & 0.914 & 0.211 {[}0.200, 0.221{]} & 0.089 \\
$\Delta$ &  &  &  &  & 0.056 & 0.074 & 0.132 & 0.061 \\ \hline
\end{tabular}%
}
\end{table}

For the final scales, several metrics were computed, which are presented in Table \ref{tab: CFA_metrics}. McDonald's omega was computed to assess the composite reliability. McDonald's omega is considered acceptable at a value greater than 0.7. The internal consistency was computed using the average variance extracted (AVE), which is considered acceptable at AVE $\geq 0.5$ and good at $\geq 0.6$. The McDonald's $\omega$ showed that both scales have excellent composite reliability. The AVE furthermore indicated that the internal consistency of the factors is very good for both scales. Finally, the factors within the scales were strongly correlated for the categorical scale and moderately correlated for the numerical scale \parencite{rottweiler2021measuring}.

\begin{table}[!ht]
\centering
\caption{Metrics for the final scale after CFA.}
\label{tab: CFA_metrics}
\resizebox{0.8\columnwidth}{!}{%
\begin{tabular}{@{}llllll@{}}
\toprule
 & Factors & Mc Donald's $\omega$ & AVE & F1 & F2 \\ \midrule
\multirow{2}{*}{Numerical} & F1 & 0.89 & 0.70 & 1 &  \\
 & F2 & 0.87 & 0.76 & 0.48 & 1 \\
\multirow{2}{*}{Categorical} & F1 & 0.92 & 0.71 & 1 &  \\
 & F2 & 0.84 & 0.71 & 0.71 & 1 \\ \bottomrule
\end{tabular}%
}
\end{table}

\subsubsection{Robustness Analysis}

\NCR{
One drawback of the current methodology is the lack of consideration of participant-wise clustering in the analysis. This is due to the limitations of the existing libraries, which do not permit the introduction of a clustering parameter in tandem with Likert-style items. To account for this limitation, a robustness analysis is conducted. For each participant, one data point is randomly retained, and the analysis is recomputed in this subsample. This is repeated across 500 random draws for both EFA and CFA, and the distribution of outcomes is reported below. The resampling led to a final dataset size of 688 (numerical EFA and CFA), 631 (categorical EFA) and 635 (categorical CFA). 

The final scales alongside the respective performance metrics and percentile ranges across draws are reported in Tables \ref{tab: numerical_scale_sens} and \ref{tab: categorical_scale_sens}. The median loadings reported across both scales differ from the full data at most by 0.02. Similarly, the median factor correlations are 0.55 and 0.65 in the subsampled data compared to 0.53 and 0.66 in the full data. The communalities are found to range from 0.57 to 0.82 (numerical) and 0.63 to 0.80 (categorical). Overall, deviations are found to be small, indicating that the participant-level clustering effect on point-estimate stability is limited. 

\begin{table}[!ht]
\centering
\footnotesize
\setlength{\tabcolsep}{4pt}
\renewcommand{\arraystretch}{1.2}
\caption{Feature Understandability Scale for Numerical Features, one-rating-per-participant robustness analysis. Values are medians across 500 random draws; brackets give the 2.5th and 97.5th percentiles of the loading across draws. $\rho$ is the polychoric item--total correlation with the item removed from the total, and $h^2$ is the communality.}
\label{tab: numerical_scale_sens}
\begin{tabularx}{\textwidth}{@{}l X c c c c@{}}
\toprule
 & & \multicolumn{2}{c}{Loading [95\% range]} & & \\ \cmidrule(lr){3-4}
N & Item & Factor 1 & Factor 2 & $\rho$ & $h^2$ \\ \midrule
\multicolumn{6}{@{}l}{\textit{Understanding \& Measurement}} \\
1 & I can understand the scale (units) of the feature. & 0.69 [0.64, 0.73] & & 0.70 & 0.57 \\
2 & I can easily understand if a given value of the feature is high or low. & 0.73 [0.69, 0.77] & & 0.68 & 0.58 \\
3 & I know what this feature measures. & 0.91 [0.88, 0.94] & & 0.76 & 0.80 \\
4 & I know what this feature represents. & 0.89 [0.85, 0.92] & & 0.78 & 0.80 \\
5 & I think that I can easily access a definition of the feature. & 0.86 [0.82, 0.89] & & 0.67 & 0.66 \\ \addlinespace
\multicolumn{6}{@{}l}{\textit{Feature-Outcome Relation}} \\
6 & In my opinion, the feature should be used to predict the outcome. & & 0.92 [0.89, 0.94] & 0.65 & 0.81 \\
7 & I think that the feature is important for the outcome. & & 0.89 [0.86, 0.91] & 0.69 & 0.82 \\
8 & I think it is fair that the feature influences the outcome. & & 0.90 [0.87, 0.92] & 0.67 & 0.81 \\
\bottomrule
\end{tabularx}
\end{table}

\begin{table}[!ht]
\centering
\footnotesize
\setlength{\tabcolsep}{4pt}
\renewcommand{\arraystretch}{1.2}
\caption{Feature Understandability Scale for Categorical Features, one-rating-per-participant robustness analysis. Values are medians across 500 random draws; brackets give the 2.5th and 97.5th percentiles of the loading across draws. $\rho$ is the polychoric item--total correlation with the item removed from the total, and $h^2$ is the communality.}
\label{tab: categorical_scale_sens}
\begin{tabularx}{\textwidth}{@{}l X c c c c@{}}
\toprule
 & & \multicolumn{2}{c}{Loading [95\% range]} & & \\ \cmidrule(lr){3-4}
N & Item & Factor 1 & Factor 2 & $\rho$ & $h^2$ \\ \midrule
\multicolumn{6}{@{}l}{\textit{Understanding \& Measurement}} \\
1 & I can easily understand how the categories were assessed. & 0.90 [0.87, 0.93] & & 0.81 & 0.80 \\
2 & I can easily understand the order of categories. & 0.91 [0.88, 0.94] & & 0.77 & 0.75 \\
3 & I think it is feasible for me to verify the specific category of the feature. & 0.70 [0.66, 0.73] & & 0.79 & 0.67 \\
4 & I understand all possible values of the categorical feature. & 0.83 [0.79, 0.86] & & 0.71 & 0.63 \\
5 & I know what this feature represents. & 0.81 [0.78, 0.85] & & 0.81 & 0.74 \\
6 & I require no support to understand the feature. & 0.78 [0.74, 0.81] & & 0.79 & 0.71 \\ \addlinespace
\multicolumn{6}{@{}l}{\textit{Feature-Outcome Relation}} \\
7 & In my opinion, the feature should be used to predict the outcome. & & 0.83 [0.79, 0.87] & 0.66 & 0.70 \\
8 & I think that the feature is important for the outcome. & & 0.81 [0.77, 0.84] & 0.64 & 0.66 \\
9 & I think it is fair that the feature influences the outcome. & & 0.88 [0.84, 0.92] & 0.69 & 0.79 \\
\bottomrule
\end{tabularx}
\end{table}

Additionally, the CFA was repeated with the same repeated sampling one-sample-per-participant approach and the results are presented in Table \ref{tab: CFA_Results_sens}. Similar to the original scale, the $\chi^2$ test remains significant, likely due to the large sample size, and the RMSEA exceeds the quoted 0.08 limit, which may be attributable to the small degrees of freedom as well as the data structure. However, the SRMR estimates, as well as the CFI and TLI, indicate good fit and diverge only minimally from the full data metrics. However, the remaining metrics are within the required ranges, indicating a good fit of the scale in the unseen data.

\renewcommand{\arraystretch}{1.2}
\begin{table}[!ht]
\centering
\caption{Results of the 2-factor CFA for the numerical and categorical feature scales across 500 random one-rating-per-participant draws. $\chi^2$, CFI, TLI and RMSEA are the scaled (mean-and-variance adjusted) statistics; SRMR is the standard estimate. ``Median'' rows give the median across draws (for RMSEA, the medians of the point estimate and of the lower and upper 90\% CI bounds); ``Range'' rows give the 2.5th--97.5th percentile of each statistic across draws.}
\label{tab: CFA_Results_sens}
\resizebox{\columnwidth}{!}{%
\begin{tabular}{lllllllll}
\hline
 & & $\chi^2$ & $\chi^2$ df & $\chi^2$ p & CFI & TLI & RMSEA {[}90\%CI{]} & SRMR \\ \hline
Numerical & Median & 123.126 & 19 & $p \leq 0.001$ & 0.989 & 0.984 & 0.089 {[}0.075, 0.105{]} & 0.035 \\
 & Range & 85.6--169.3 & 19 & $p \leq 0.001$ & 0.985--0.993 & 0.977--0.990 & 0.071--0.107 & 0.028--0.041 \\
Categorical & Median & 134.884 & 26 & $p \leq 0.001$ & 0.991 & 0.988 & 0.081 {[}0.068, 0.095{]} & 0.030 \\
 & Range & 104.8--175.9 & 26 & $p \leq 0.001$ & 0.988--0.994 & 0.983--0.991 & 0.069--0.095 & 0.026--0.035 \\ \hline
\end{tabular}%
}
\end{table}
}

The full scoring guidance alongside worked examples is available on GitHub\footnote{\url{https://github.com/ncrossberg/Feature-Understandability-Scale}}.

\section{Conclusion} \label{sec: Discussion}

The \textit{Feature Understandability Scale} measures participants' perceived understanding of features in a tabular dataset. The ultimate goal of the scale was to quantify understandability at the feature level for applications in explainable ML and interpretability by design. This differs significantly from previous scales in the field, shifting the measurement unit from the explanation to the feature level and moving the focus point from post-hoc to ante-hoc. The initial scale was drafted based on the literature and refined through expert interviews, focus groups and a pilot study. Following the steps laid out by \textcite{carpenter2018ten}, an exploratory factor analysis was then conducted. Two two-factor scales, one for numerical items and one for categorical items, were extracted. To confirm this structure, a confirmatory factor analysis was conducted on a separate dataset. The two-factor structure was confirmed, showing significant improvement over a one-factor solution. The factors of both scales were named according to the same logic, with the first factor, 'Understanding and Measurement', containing items that assessed whether users understood the concept underlying the feature, as well as how the value of the feature was measured. The second factor, `Feature-Outcome Relation', contains the same items for both scales, encompassing how the participant perceives the feature as a predictor for the outcome. It is important to note that the second factor captures users' perceptions of whether a feature is a sensible, relevant input for the model's prediction. It does not capture, and should not be interpreted as, a judgment of causal relationships between the feature and the real-world outcome. The final scale showed good composite reliability and internal consistency. It is, however, important to note that the current analysis did not account for participant-wise clustering. \NCR{While a robustness analysis showed stable factor loadings and model fit, it did not account for impacts on standard errors, test statistics or reliability estimates. As such, the proposed structure should be further validated in future research to ensure that the participant-wise clustering effect does not significantly impact its structure and reliability.}

This work developed a scale to measure the understandability of features, marking an important first step to a new method for interpretability by design. The quantification of understandability at the feature-level permits an ante-hoc approach to xAI, which has not been previously considered. While previous works, including \textcite{holzinger2020measuring, vilone2023development, brooke1996sus}, proposed scales to measure the quality of explanations after their generation, their methods are insufficient to improve their quality, operating at the descriptive rather than the active level. \NCR{The current scale provides the unique advantage that, through the co-optimisation of understandability and accuracy in the machine learning pipeline, explanation quality can be targeted as a design consideration rather than a post-hoc component. While further validation of the final scale structure is necessary to account for clustering effects, initial work by \cite{rossberg2026integrating} showed that the integration of the scale may permit the creation of more understandable explanations.} Additionally, previous scales often took a 'quick and dirty' approach to assessing explanation quality, whereas the current work underwent psychometric validation. The developed scale holds several important implications. On the theoretical end, the quantification of understandability is a novel contribution and may find applications in psychometrics and computer science outside of the xAI domain. In practice, the proposed scale permits the measurement of understandability scores, which can be implemented into the machine learning pipeline and used in tandem with existing explainability methods to prioritise understandable features during model training. In addition to the potential for co-optimisation, the scale offers practical advantages, including the potential to improve auditing by assessing the understandability of a model's important features, and a shifting focus towards human-centred AI design.

\subsection{Future Works}

Given the multi-disciplinary nature of this work, several approaches to future research should be considered. From a psychometric perspective, validation of the scale structure in new samples should be prioritised to elucidate whether the structure is replicable. Additionally, it may be of interest to stratify the sample by gender, ethnicity or level of education to test whether the current scale structure is an artefact of sample fluctuations. While efforts were made to mitigate these effects in the current study through pseudo-random sampling and a large sample size, they could not be entirely negated. Furthermore, assessing measurement invariance across relevant demographics (e.g., level of domain-based education, confidence, etc.) would be beneficial. \NCR{Additionally, it would be important to repeat the scale validation in a novel dataset without repeated sampling between participants or otherwise accounting for the potential clustering effect. While this study implemented a robustness analysis to test the effect of the repeated-measures design, further validation in a new sample would be beneficial to test the identified structure.} This work was further novel in proposing a potential interaction between participants and features, and future research should elucidate its nature. While the scores on the FUS are expected to be treated as random effects at the participant level, this should be the subject of future work. In addition to scale replication, translation to other languages and data topics may be of interest. While translation into other languages would permit insight into whether the scale structure is an artefact of the English language, translation into other domains, such as measuring understanding in a language acquisition context, could open the door to broader adaptation. 

From an ML perspective, the \textit{Feature Understandability Scale} marks the first step towards integrating `understanding' scores into the ML pipeline, enabling the prioritisation of understandable features during training. This is advantageous, as the scale functions on an a priori basis and thus complements rather than replaces existing xAI evaluation methods. By integrating understandability scores into the machine learning training pipeline, understandability and accuracy can be co-optimised prior to the creation of explanations post-hoc. To test whether the scale can improve explanations, future research should aim to integrate it into the ML pipeline, favouring variables with high understandability scores during the explanation process and evaluating the quality of the explanations through a user study. This assessment would also permit the assessment of the potential accuracy-understandability trade-off, testing whether the favouring of understandable features significantly affects accuracy levels across different domains. It may furthermore be of interest to assess where in the ML workflow scores are best integrated to improve explanations. While integrating the scores before or in the ML algorithm would permit an interpretability-by-design solution, in which more understandable features are favoured during the training process, this may present a trade-off with accuracy, as the more understandable features are not necessarily the most informative ones. On the other hand, the scores could serve as a filter method after ML training and before creating an explanation, where relevant features are refined and only understandable features are included in the final explanation. While this would negate the accuracy-explainability trade-off, it would reduce the authenticity of the explanation, as not all relevant features may be included. \NCR{ Another possible application would be in the context of counterfactual explanations, where the scale could be used to drive the selection of features, the change of which would positively affect the output.} In addition, it should be assessed whether the use of understandable features consistently leads to the creation of more understandable explanations, or whether other moderators, such as domain, education level, or explanation length, need to be considered. 

\subsection{Limitations}
Despite best efforts, certain limitations of this research need to be acknowledged. First, while pseudo-random sampling was employed through the platform `Prolific', a true-random sample could not be collected; hence, there remains a chance that the current sample structure is an artefact of sample demographics. Additionally, the sample remains limited to English-speaking, Western populations and Prolific users and generalisability outside this sample, as well as the three domains within which it was validated, is not guaranteed.

\NCR{The scale furthermore relies on self-reported rather than objective understanding and hence may be affected by factors including confidence and domain knowledge. The effects of the latter within the collected sample were not explored, although this should be somewhat mitigated by the random shuffling of features and domains, as well as by the division of participants' responses between categorical and numerical data. However, this is an important aspect to consider in future research, as understanding is not expertise-independent and may be moderated by domain understanding, and a multi-level modelling approach should be explored.}

\NCR{An important limitation of the current scale construction methodology is the repeated-measures design, taking multiple feature ratings from each participant, which is not accounted for during data evaluation. The results of the robustness analysis showed stable factor loadings and model fit but addressed only the participant-level clustering, while retaining the same participants in every draw. As such, it does not account for potential effects on standard errors, test statistics or reliability estimates. Future research should aim to verify the scale's structure in a sample without repeated measures or else integrate a clustering component directly into the EFA and CFA methodology.}

It is also important to note that while the final large-sample validation was based on random sampling on Prolific, the previous refinement steps relied on recruiting experts and laypeople from the researchers' networks and on snowball sampling during the pilot study. While the aim of these steps was not statistical generalisability but the introduction of expert feedback and refinement of item phrasing, it is important to note that such qualitative refinement may introduce some bias. 

Furthermore, no test-retest reliability was computed, and longitudinal validation of the study may be of interest in the future. One potential confound of the current study is the lack of multilevel analysis. While participants were randomly assigned to domains and feature types (categorical and numerical), the potential for participant and domain effects cannot be dismissed.

Finally, despite following the recommendations for best practices set out by \textcite{carpenter2018ten}, a possibility remains that items were missed during the initial creation of the scale, and hence the final scale may be incomplete. As such, future research should aim to replicate both the final scale structure and the scale construction process to ascertain whether any relevant considerations were not included in the current research. In addition to the internal limitations of the scale construction, the work itself is limited to tabular data only and cannot be applied to other data types.

The study is the first step toward an explainability-by-design approach that quantifies the user's domain knowledge and enables its integration into the ML workflow. This enables the co-optimisation of accuracy and explainability for tabular data, producing explanations that are understandable by default and enhancing the transparency and auditability of ML models. While future research should aim to replicate and integrate the existing scale, this article marks an important first step in a new approach to explainability-by-design solutions. 

\section*{Funding}
This work was conducted with the financial support of Research Ireland - Taighde Éireann, under Grant Nos. 18/CRT/6223 and 12/RC/2289-P2, the latter being co-funded under the European Regional Development Fund. For the purpose of Open Access, the author has applied a CC BY public copyright licence to any Author Accepted Manuscript version arising from this submission.

\section*{Declaration of the use of AI}
During the preparation of this work, the authors used 'Grammarly' in order to correct and improve the writing of this manuscript and 'Claude' (Anthropic; Sonnet 5.5) to adjust the formatting of references, support code production, draft the scoring guidance, and transcribe output into tables. After using these tools, the authors reviewed and edited the content as needed and took full responsibility for the content of the published article. No generative AI was used to create or edit the graphical abstract.

\section{Data Availability Statement}
All relevant data and code can be found at \url{https://github.com/ncrossberg/Feature-Understandability-Scale}. The data is completely anonymised and shared in line with ethics guidelines.

\section{Declaration of competing interests}
The authors declare that they have no known competing financial interests or personal relationships that could have appeared to influence the work reported in this paper.

\newpage
\newpage

\printbibliography

\end{document}